\documentclass[superscriptaddress,prb,twocolumn,amsmath,amssymb]{revtex4}

\usepackage{graphicx}
\usepackage{color}
\usepackage[usenames,dvipsnames]{xcolor}
\usepackage{eurosym}
\usepackage{dcolumn}
\usepackage{bm}
\usepackage{subfigure}
\usepackage[final]{pdfpages}
\usepackage{setspace}
\usepackage{pdfpages}
\usepackage{ulem}
\usepackage[draft]{pgf}
\graphicspath{{./images/}}

\newcommand{\ket}[1]{\left | #1 \right \rangle}
\newcommand{\bra}[1]{\left \langle #1 \right |}

\newcommand{\jm}[1] {\textcolor{black}{#1}}
\newcommand{\jmob}[1] {\textcolor{black}{#1}}
\newcommand{\pete}[1] {\textcolor{black}{#1}}
\newcommand{\durk}[1]{\textcolor{black}{#1}}

\begin{document}

\title{\jmob{Practical quantum metrology}}

\author{Jonathan C. F. Matthews\footnote[1]{These authors contributed equally}$^{,\,}$\footnote[2]{Email: Jonathan.Matthews@Bristol.ac.uk}$^{,\,1}$, 
Xiao-Qi Zhou$^*$$^{,\,1}$,
Hugo Cable$^*$\footnote[3]{Email: cqthvc@nus.edu.sg}$^{,\,2}$, 
Peter J. Shadbolt$^*$$^{,\,1}$, 
Dylan J. Saunders$^3$, 
Gabriel A. Durkin$^4$,
Geoff J. Pryde\footnote[4]{Email: G.Pryde@Griffith.edu.au}$^{,\,3}$,
Jeremy L. O'Brien\footnote[5]{Email: Jeremy.OBrien@bristol.ac.uk}$^{,\,}$}
\affiliation
{
Centre for Quantum Photonics, H. H. Wills Physics Laboratory
and Department of Electrical and Electronic Engineering, University
of Bristol, Merchant Venturers Building, Woodland Road,
Bristol BS8 1UB, UK.\\
$^\textrm{\textit{2 }}$Centre for Quantum Technologies, National University of Singapore\\
$^\textrm{\textit{3 }}$Centre for Quantum Dynamics and Centre for Quantum Computation and
Communication Technology, Griffith University, Brisbane 4111, Australia\\
$^\textrm{\textit{4 }}$Google/NASA QuAIL, NASA Ames Research Center, Moffett Field, California 94035, USA 
}

\date{\today}

\begin{abstract}
Quantum metrology research promises approaches to build new sensors that achieve the ultimate level of precision measurement and perform fundamentally better than modern sensors. Practical schemes that tolerate realistic fabrication imperfections and environmental noise are required in order to realise quantum-enhanced sensors and to enable their real-world application. We have demonstrated the key enabling principles of a practical, loss-tolerant approach to photonic quantum metrology designed to harness all multi-photon components in spontaneous parametric downconversion---a method for generating multiple photons that we show requires no further fundamental state engineering for use in practical quantum metrology. We observe a quantum advantage of 28\% in precision measurement of optical phase using the four-photon detection component of this scheme, despite 83\% system loss. This opens the way to new quantum sensors based on current quantum-optical capabilities.
\end{abstract}

\maketitle

\jm{Quantum shot noise represents a hard limit for the precision of all modern sensors that do not harness 
non-classical
resources. Photonic quantum metrology~\cite{gi-nphot-5-222} promises to surpass the shot-noise limit (SNL) by using quantum states of light that exhibit entanglement~\cite{do-contphys-49125}, discord~\cite{mo-prx-1-021022} or squeezing~\cite{go-nphys-4-472} to suppress statistical fluctuation. However, there is no existing sensor that routinely employs these resources to obtain sub-SNL performance. Critically, this is due to unavoidable optical loss \jmob{that severely hinders} 
quantum advantage}\pete{,}
while schemes aimed at tolerating loss for sub-SNL performance have previously required fixed photon-number states. This is currently infeasible since the only implemented approach to access fixed photon-number $N > 2$ is based on post-selection from the whole multi-photon Spontaneous Parametric 
Downconversion (SPDC) state --- this is random, produces unwanted photon-number states and ultimately demands complex 
heralding techniques to filter the light. 

Here we demonstrate the key enabling principles of a practical 
loss-tolerant scheme for sub-shot-noise interfereometry~\cite{ca-prl-105-013603}, that uses the full multi-photon state naturally occurring in type-II SPDC. \jm{To this end, we} \jmob{overhaul} \jm{the theoretical analysis of \jmob{the original proposal~\cite{ca-prl-105-013603}} into a form suited for arbitrary number counting methods,}
\pete{such as the multiplexed detection scheme used here.}
\jm{
We measure the four-fold detection events arising from all SPDC contributions of four or more photons, and \jmob{observe} a quantum advantage of 28\% in the mean squared error \jmob{of optical phase estimation} in the presence of 83\% combined circuit and detector loss. This scheme provides a simple and practical method for loss-tolerant quantum metrology using existing technologies.
}

Photon-counting experiments investigating the principles of multi-photon interference in an interferometer \cite{ra-prl-65-1348} were followed by a series of interference experiments with increasing photon number for quantum metrology \cite{mi-nat-429-161,wa-nat-429-158,na-sci-316-726}.
The goal of 
quantum metrology is to estimate or detect a\jmob{n optical}  phase~$\phi${---that can \jmob{map directly} to distance, birefringence, \jm{angle,} sample concentration \jm{etc.}---}with precision beyond the \jm{SNL} ($\Delta \phi \sim 1/\sqrt{N}$) in the low-photon-flux regime. {Optimising photon flux to gain maximal information can be useful to minimise detrimental effects from probe light in biological sensing, for example.} \jmob{A much sought after objective}
has been to engineer ``NOON'' states~\cite{do-contphys-49125}---path-entangled state{s} of $N$ photons across two modes $\frac{1}{\sqrt{2}}(\ket{N}\ket{0}+\ket{0}\ket{N})${---that} offer both super-resolution {(}N-fold decrease in fringe {period)} and super-sensitivity {(enhanced precision towards the Heisenberg limit\jmob{---$\Delta \phi \sim 1/{N}$})}. The current record in size of NOON-like states is five photons using postselection \cite{af-sci-328-879}, and four photons using \pete{ancillary-photon detection}~\cite{ma-prl-107-163602}.

\pete{Exact} 
generation of higher-\jmob{photon-number}
NOON states using passive linear optics has exponentially-\jmob{increasing} resource requirements\jmob{; o}ne solution is to use feed-forward to efficiently generate NOON states~\cite{ca-prl-99-163604}
\pete{which}
requires \jmob{much of} the same capability as full-scale linear-optical quantum computing~\cite{kn-nat-409-46}. An even more \jmob{serious} problem is that
large NOON states perform worse than the shot-noise limit given any realistic loss, since reduction in precision is amplified by the photon number~\cite{do-contphys-49125}. Loss will be present in any practical scenario\jmob{, ranging from the use of non-unit efficiency detectors to \pete{absorbance in} bio-sensing \cite{cr-apl-100-233704,ta-nphoton-7-229}.} 
Consequently, considerable theoretical effort has been devoted to the development of schemes that minimise the detrimental effects of photon loss. 
\durk{Revised scaling laws of precision with photon flux have been derived\cite{kn-pra-83-021804}, along with optimized superposition states of fixed photon number, numerically for small photon number\cite{le-pra-80-063803, do-pra-80-013825} and analytically for large\cite{kn-pra-83-021804}.}

\begin{figure}[t!]
\centering
\includegraphics[width=\columnwidth]{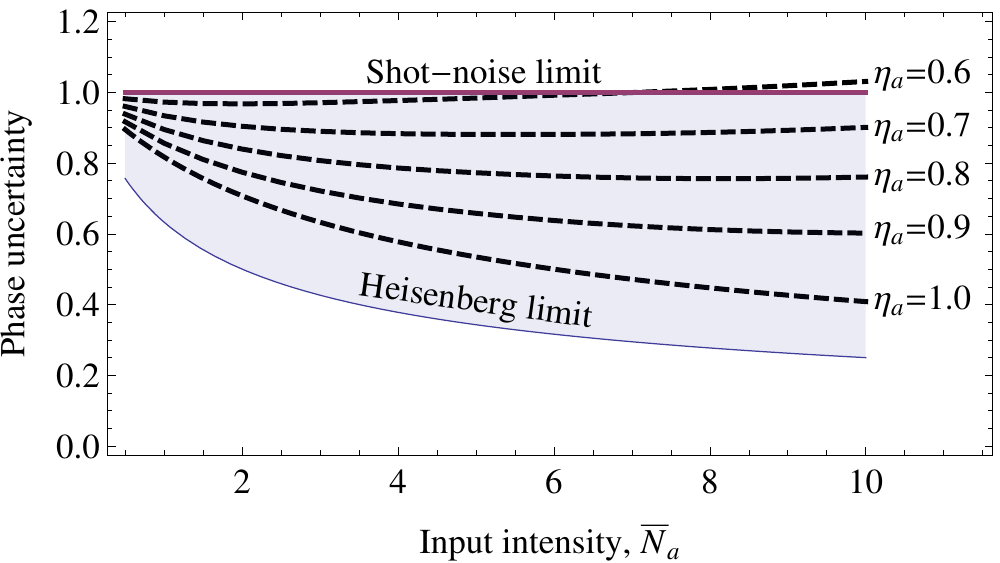}
\vspace{-0.7cm}
\caption{
\textbf{Performance of photon counting and type-II SPDC in the presence of loss $(1-\eta_a)$, without heralding or postselection.
}
{
The uncertainty of estimating or detecting an unknown phase $\phi$ is defined as $\Delta \phi \sqrt{\eta \bar{N}_a}$, renormalised by the average intensity $\eta \bar{N}_a$ in the
interferometer---equivalently one can consider the available statistical information defined as Fisher information per photon. We compare the scheme with the shot-noise limit achieved using perfect coherent laser light, and the Heisenberg limit,
$1/\sqrt{\langle\hat{N}^2_a\rangle}$ (relevant to cases where the total photon 
number fluctuates~\cite{ho-pra-79-033822}).  Loss is balanced in the sensing interferometer and
$\bar{N_a}$ 
is the average number of photons entering the interferometer, before loss.
}
}
\label{ComparisonFig1}
\end{figure}

To date, all demonstrations aimed at developing quantum technology with photon counting are designed to operate with a deterministically generated \jmob{fixed} number of photons --- this includes approaches to loss tolerant quantum metrology, such as Holland \& Burnett states~\cite{ho-prl-71-1355,xi-natphot-5-43}. The only system that has demonstrated quantum interference of more than 2 photons is SPDC --- a nonlinear process that generates a coherent superposition of correlated photon-number states. To perform experiments with $n$ photons, post-selection is employed to ignore components of fewer photons ($<n$), while terms associated with higher photon number ($>n$) are treated as noise. This is 
particularly 
problematic for quantum metrology, where all photons passing through the sample need to be accounted for and unwanted terms are detrimental to measurement precision. 

Here we adopt an alternative approach designed to achieve sub-SNL performance using the entire four-mode multi-photon entangled state naturally generated in type-II SPDC \cite{kw-prl-75-4337,la-nat-412-887,si-prl-91-053601}. This state {is a superposition of all photon-number 
singlet states and }can \jmob{surprisingly} achieve Heisenberg scaling in the absence of decoherence~\cite{ca-prl-105-013603}, {in a }similar {manner} to NOON states. More importantly, this state surpasses the shot-noise limit despite a realistic level of loss {that would} \jmob{otherwise} preclude any quantum advantage {when} using NOON states.
Fig.~\ref{ComparisonFig1} illustrates the sub-SNL performance of photon counting on the Type-II SPDC scheme in the presence of loss.
\jm{Intuition for the loss tolerance in this scheme 
can be gained by considering 
the effect of 
losing a single photon from one of the modes: each singlet component transforms into a state that closely approximates another singlet of lower photon number~\cite{la-nat-412-887}.
}
\jm{We observe sub-SNL phase sensitivity in the four-photon coincidence detection subspace of our experiment, i.e. using all four-photon detection events due to singlet components with $N \geq 4$ photons. We do not post-select zero loss or assume a fixed photon number in our theoretical analysis. 
}
\jmob{This supports
the loss-tolerance 
expected from
detecting $N>4$ from any higher photon number components of type-II SPDC~\cite{ca-prl-105-013603}.}

\begin{figure}
\includegraphics[width=\columnwidth]{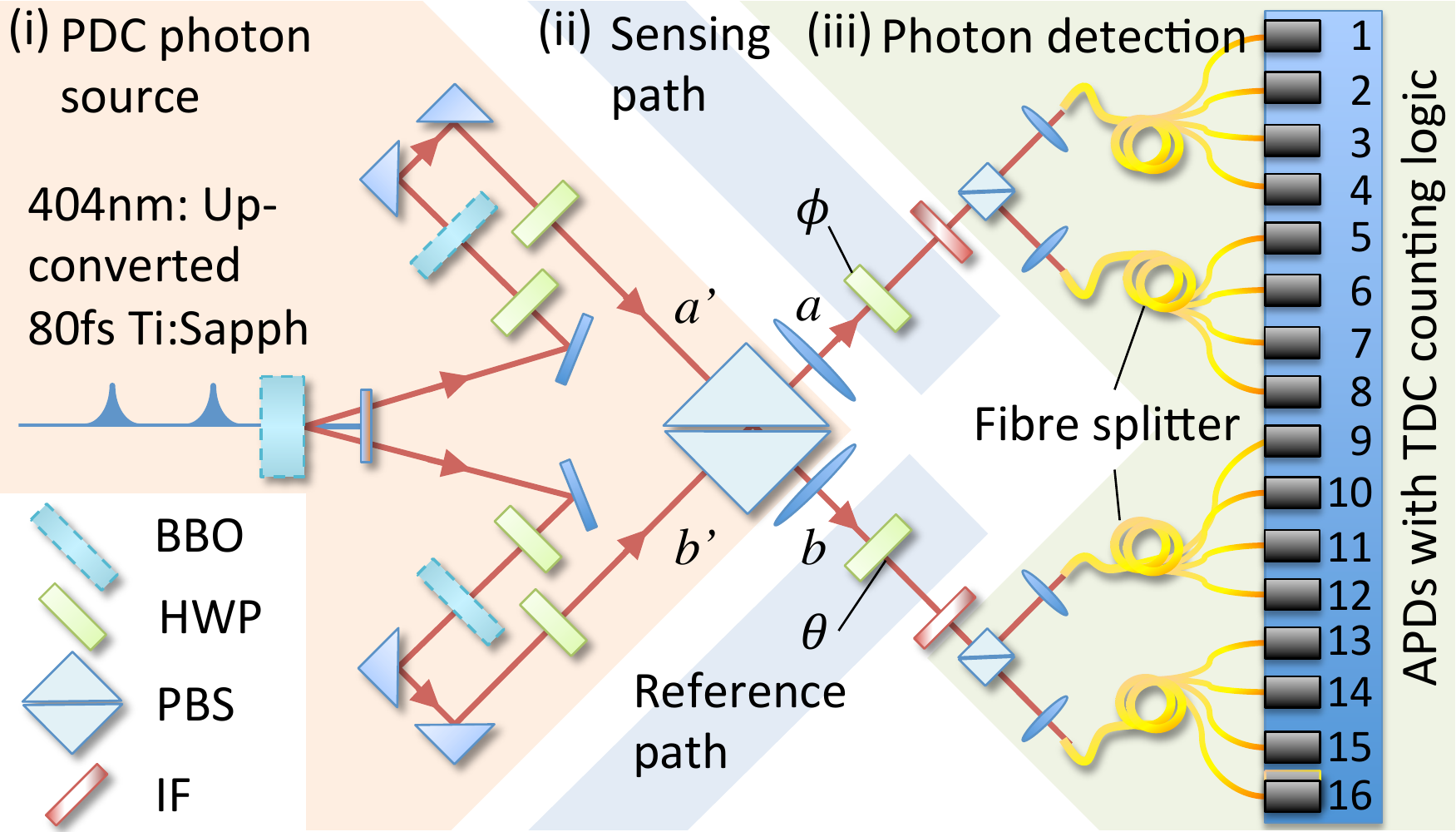}
\vspace{-0.7cm}
\caption{\textbf{Schematic of the experimental setup.}  The probe state $\ket{PDC}$ of polarisation-entangled photons
is generated in a 2mm thick $\chi^2$ nonlinear Barium Borate (BBO) crystal phase-matched for non-colinear type-II SPDC\cite{kw-prl-75-4337}.
\jm{We use a Polarising Beamsplitter (PBS) to remove spectral-path correlation\cite{ki-pra-67-010301} to ideally generate the desired state $\ket{PDC}$ across modes $(a_h, a_v, b_h, b_v)$ and spectrally filtered with Interference Filters (IF) (see 
\jmob{Appendix}). The phase parameter~$\phi$ 
is implemented using a half-waveplate (HWP).}
}
\label{setup1}
\end{figure}

Our demonstration (see Fig.~\ref{setup1}) can be treated in \jmob{the} three stages (i) the source, (ii) \jm{unitary rotation with an unknown parameter $\phi$ to be estimated on the sensing path $a$} 
and (iii) the photon-counting measurement. (i) The source is based on a non-colinear type-II SPDC~\cite{kw-prl-75-4337,la-nat-412-887}, that generates \pete{entanglement} across four modes\pete{---two spatial paths ($a$, $b$) and two polarizations ($h$, $v$) (see Appendix).}
In the ideal case, for which all decoherence is neglected, 
\jm{the state generated is} the superposition of photon-number states
\begin{eqnarray}
\ket{PDC}  \! \propto \! \! \sum_{n=0}^\infty \left(\tanh \tau\right)^n\!\!\left[\! \sum_{m=0}^n (-\!1)^m \! \ket{n\!-\!m,m,m,n\!-\!m} \!\right]
\label{SPDCsinglet}
\end{eqnarray}
where $\tau$ is an interaction parameter that corresponds to the parametric gain, and the modes are listed in order $(a_h, a_v, b_h, b_v$). Note that we have omitted normalisation. This state has the property that each term indexed by $n$ corresponds to an entangled state having a total of $2n$ photons, and maps onto the singlet state that represents two spin-$n/2$ systems in the Schwinger representation \cite{Sakurai}.
When $\tau$ is small, $\ket{PDC}$ is dominated by the $n=1$ term which enables post-selection of the two-photon entangled state $\frac{1}{\sqrt{2}}\left(\ket{H}_a\ket{V}_b - \ket{V}_a\ket{H}_b\right)$\cite{kw-prl-75-4337}.
For larger $\tau$, the photon intensity grows as\cite{la-nat-412-887} $\sim2\sinh ^2 \tau$. The symmetry and correlation properties of $\ket{PDC}$ have been the subject of several investigations, with experimental evidence reported for entanglement between $\sim100$ photons~\cite{ei-prl-93-193901}, \jmob{with} possible applications \jmob{proposed outside of metrology}~\cite{la-nat-412-887,ra-pra-80-040302}.

\pete{(ii) The rotation we consider is
\begin{eqnarray}
U_{a'}(\phi)\doteq\left(
\begin{array}{cc}
  \cos (\phi/2 ) &  \sin (\phi/2 ) \\
  \sin (\phi/2 ) & - \cos (\phi/2 ) \\
\end{array}
\right),
\end{eqnarray}
where $\phi$ is the parameter we wish to estimate with quantum-enhanced precision. This operator maps exactly to rotations of any two-level quantum system,
including
the relative phase shift in an interferometer. We implement 
$U_{a}(\phi)$
using a 
half-waveplate in the sensing path $a$, operating on 
modes $\{a_h, a_v\}$,
and for which $\phi$ is four times the waveplate's rotation angle.
}

(iii)~Finally, photons in each of the four modes $\left(a_h,a_v,b_h,b_v\right)$ are detected with number-resolving photodetection---the
original proposal \cite{ca-prl-105-013603} assumed 
\durk{fully photon-number resolving}
detectors 
\jmob{that}
implement projections onto all Fock states. 
\pete{We developed an efficient 
number resolving multiplexed detection system using readily available components including four $1\times 4$ optical fibre splitters; sixteen Avalanche Photodiode single photon counting modules (APDs) and a novel sixteen-channel real-time coincidence counting system that records all possible combinations of multi-photon detection events occurring coincidentally across the sixteen APDs (see Appendix).}

A powerful method to simplify
calculating measurement outcome probabilities
for our experiment is to use the Positive-Operator-Valued Measurements~(POVM) formalism{~\cite{Peres}}. 
All photon-counting operations correspond to POVM elements $E_r$, that are diagonal in the Fock state basis $\{\ket{c}\}$:  $E_r = \sum_{c=0}^\infty w_r\left( c\right) \ket{c}\!\bra{c}$, where $r$ and $c$ denote the detection pattern and the photon number respectively. The weights $w_r(c)$ are non-negative and satisfy 
$\sum_r w_r(c)=1$. The probability of $r$ detection events is 
given by $P_r = tr(\rho \hat{E}_r)$, where $\rho$ is the density matrix of any state input to the measurement setup.
For the perfectly number-resolving case, the only non-zero POVM weight is when $c=r$ and $w_r(r)=1$. However, with 
multiplexed detection, 
all weights $w_r(c)$ with $c\geq r$ can be non-zero. For example, for a two-photon state incident on one of our multiplexed \jmob{detectors,}
there is a probability of $1/4$ that both photons go to the same APD causing one detection event~($w_1(2)=1/4$) and a probability of $3/4$ for two detection events~($w_2(2)=3/4$). The entire table of the POVM weights for our multiplexed system are explained in the Appendix (see also Ref.~\onlinecite{sp-pra-85-023820}).

Multiplexed detector POVMs are applied to each of the modes $a_h$, $a_v$, $b_h$ and $b_v$ to compute the probability for a detection outcome $\mathbf{r} = (r_{ah},r_{av},r_{bh},r_{bv})$, given a phase rotation $\phi$:
\begin{eqnarray}
P_\mathbf{r}\!\left(\phi \right) =\!\!\!\!\!\!\!\!\!\!\!\!\!\!\!\!\sum_{c_{ah},c_{av},c_{bh},c_{bv}\geq 0}\!\!\!\!\!\!\!\!\!\!\!\!\!\!\! w_{r_{{ah}}}\!(c_{ah}) w_{r_{{av}}}\!\left( c_{av}\right)  w_{r_{{bh}}}\!\left( c_{bh}\right)  w_{r_{{bv}}}\!\!\left( c_{bv}\right)p_{\mathbf{c}}\!\left(\!\phi\right)
\nonumber\\
\label{POVMequation}
\end{eqnarray}
where
$\mathbf{c}=\left(c_{ah},c_{av},c_{bh},c_{bv}\right)$ {is the photon number for each mode}
and $p_{\mathbf{c}}\!\left(\!\phi\right)$
corresponds to the probability
for a measurement outcome of a perfect projection $\ket{c_{ah},c_{av},c_{bh},c_{bv}}\!\bra{c_{ah},c_{av},c_{bh},c_{bv}}$. 
\jm{From  Eq.~(\ref{SPDCsinglet}), rotation on modes $a_h$ and $a_v$ yields the probability to detect $\mathbf{c}$ according to}
\begin{eqnarray}
p_{\mathbf{c}}=\delta_{c_a,c_b}
\frac{\tanh ^{2 c_{a}}\tau}{\cosh ^{4}\tau }
\,\, \!|\!\bra{c_{ah},c_{av}} U(\phi) \ket{c_{bv},c_{bh}}\!|^2
\end{eqnarray}
where photon number for the two paths are denoted by $c_a=c_{ah}+c_{av}$ and $c_b=c_{bh}+c_{bv}$, and
\jm{where the Wigner-d matrix element
$d_{m',m}^{j}\left( \phi \right)^2\!\!=\!|\!\bra{j\!+\!m',j\!-\!m'} U(\phi) \ket{j\!+\!m,j\!-\!m}\!|^2$
describes the rotation amplitudes on two separate modes populated by number states~\cite{Sakurai}, and is conveniently represented as a cosine Fourier series~\cite{ca-prl-105-013603}.}

Losses are straightforward to incorporate into this formalism.  Our model assumes all detectors have the same efficiency and there is no polarisation-dependent loss in our setup, therefore all loss that can arise in our setup commutes with $U_{a}(\phi)$.
We incorporate the total circuit and detector efficiency ($\eta$)
into the POVM elements via a simple adjustment of 
$w_r(c)$ (see Appendix).

\begin{figure}[t!]
\includegraphics[width=
\columnwidth
]
{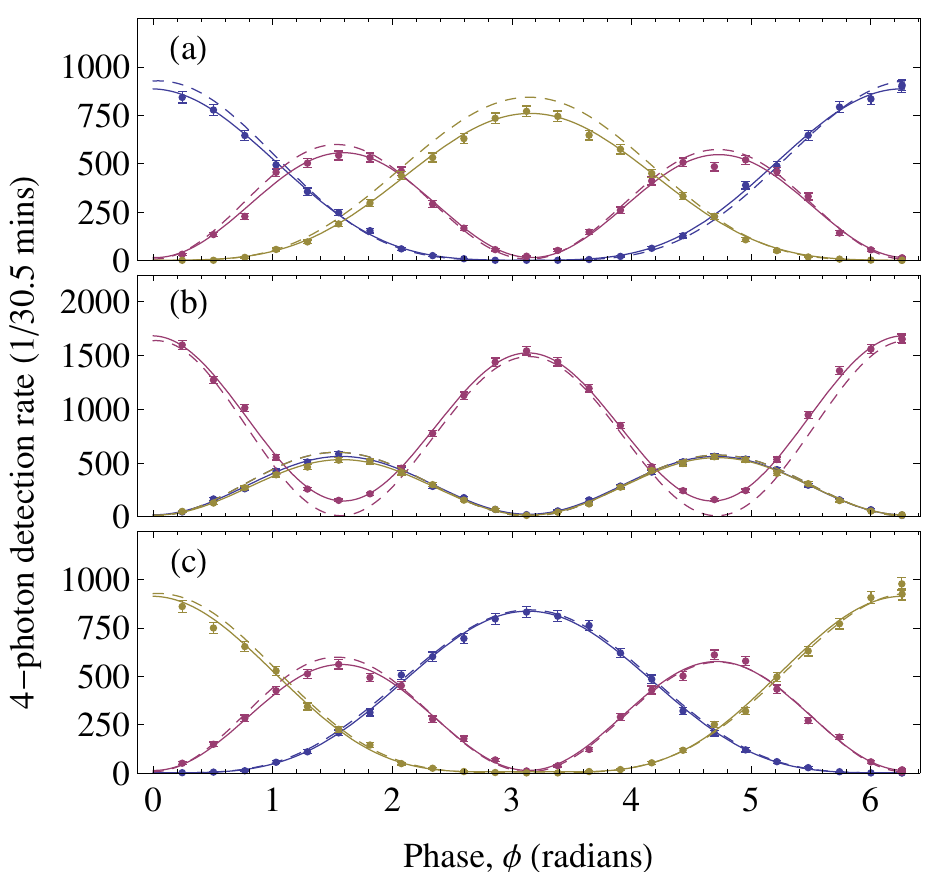}
\vspace{-0.7cm}
\caption{
\textbf{
Four-photon interference fringes of $\ket{PDC}$.
} 
The detection patterns $\mathbf{r}$ (circular points)
are: (a) 2002 (blue), 2011 (red) and 2020 (gold); (b) 1102 (blue), 1111 (red) and 1120 (gold) and (c) 0202 (blue), 0211 (red) and 0220 (gold). Error bars are computed assuming Poisson-distributed noise on 
detection statistics. Curves of best-fit (solid lines) are computed 
using functions derived from theory---$\sum_{s=0}^{2} C_s \cos(s(\phi + \phi_o))$,
for free parameters $C_s$ and $\phi_o$.
Theoretical 
distributions $P\!\left( \mathbf{r},\phi \right)$ (dashed lines) are computed for each $\mathbf{r}$ with characterised parameters $\tau=0.061$, $\eta_{a}=0.23$ $\eta_{b}=0.12$.
}
\label{9probspatterns}
\end{figure}

We plot in Fig.~\ref{9probspatterns} all nine possible four-photon detection patterns
$\mathbf{r}$ of two photons in the reference path and two photons in the sensing path as a function of $\phi$, measured simultaneously by the setup in Fig.~\ref{setup1}. For comparison, we plot this data together with theoretical curves $P(\mathbf{r},\phi)$, \jm{normalising} to the total counts collected at each $\phi$. 
These theoretical curves use the measured experimental parameters of
$\tau=0.061$,
and lumped collection/detection efficiencies of
$\eta_{a}=0.23$ and $\eta_{b}=0.12$
in the sensing and reference paths respectively 
(a geometric average of $83.2\%$ loss), assuming otherwise perfect $U_{a}(\phi)$ and photon interference.  
The asymmetry in $\eta_a$ and $\eta_b$ arises from the different spectral width of the extraordinary and ordinary light on respectively paths $a$ and $b$, passing through identical spectral filtering~\cite{gr-pra-56-1627}.
\durk{The setup is robust to this since the state symmetry is preserved despite 
$\eta_a \neq \eta_b$,
provided
loss is polarisation insensitive \cite{la-nat-412-887}.}
\jm{From the 
data 
presented in Fig.~\ref{9probspatterns}, we extract the probability distributions $p_i(\phi)$ as least-squares fits from each data set, and normalise such that \mbox{$\sum_i p_i(\phi)=1$}.} 

\begin{figure}
\includegraphics[width=\columnwidth]{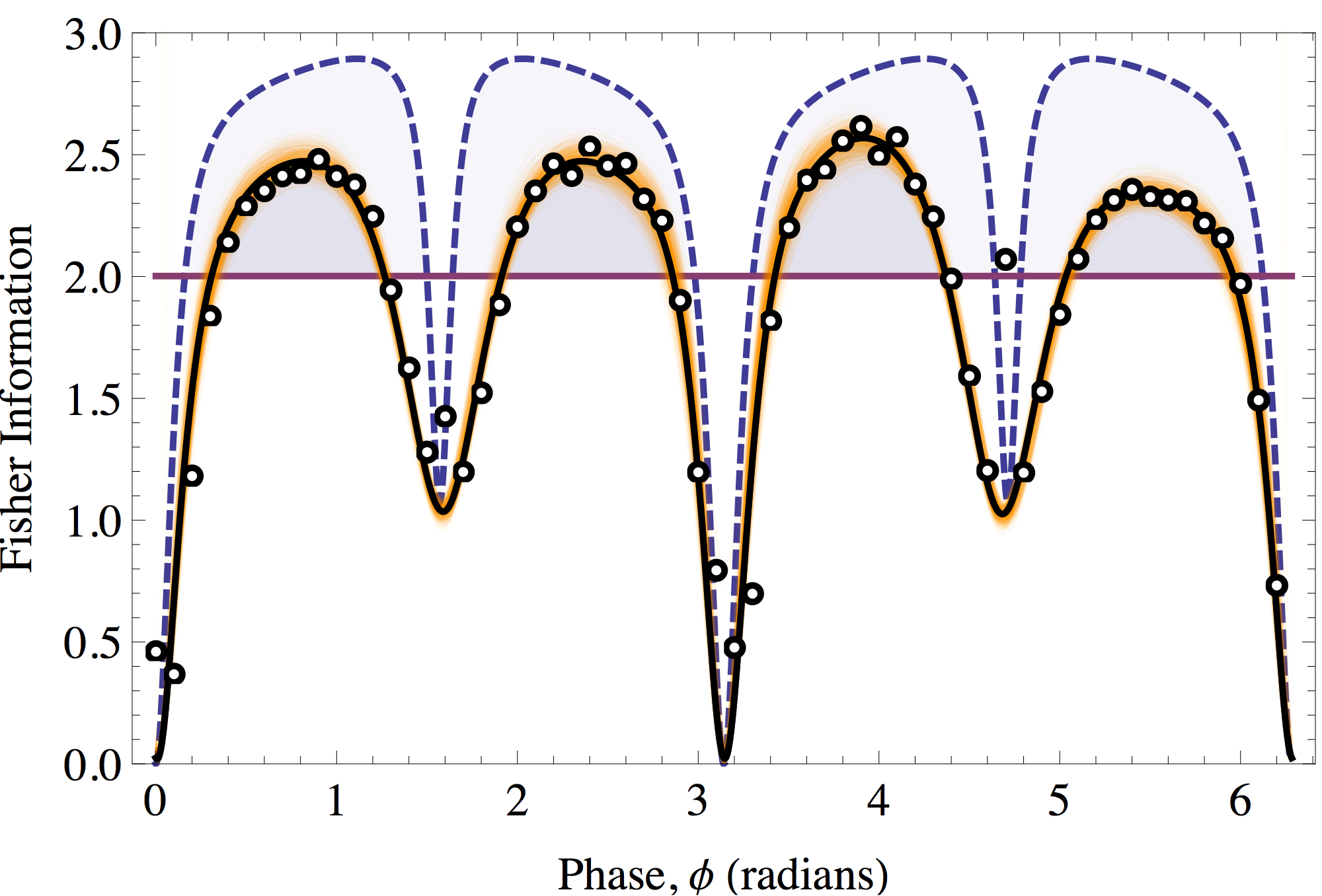}
\vspace{-0.7cm}
\caption{
\textbf{Fisher information extracted $\ket{PDC}$ interference fringes}
\textit{Solid black line:} Total fisher information $\mathcal{I}(\phi)$ for the fitted probability distributions $p_i(\phi)$ from Fig.~\ref{9probspatterns}. \textit{Orange lines:} 1,000 iterations of a Monte-Carlo simulation of $\mathcal{I}(\phi)$, assuming possonian noise on the raw photon counts. \textit{Blue dashed line:} Theoretical Fisher information for our setup, computed with the parameters
$\tau=0.061$, $\eta_{a}=0.23$ $\eta_{b}=0.12$. \textit{Purple line:} The shot-noise limit. \textit{Shaded regions:} depicts where the scheme theoretically and experimentally displays quantum advantage. 
\textit{Circles:} Fisher information $\mathcal{I}_{\rm ML}$ computed from maximum-likelihood estimates of $\phi$.
}
\label{FisherMonteML}
\end{figure}

\jmob{
Statistical information about $\phi$ can be extracted from the frequencies of each output detection pattern and quantified using Fisher information $\mathcal{I}(\phi)$~\cite{br-jopa-25-3813}. The importance of Fisher information lies in the Cram\'{e}r-Rao bound, which states that any unbiased statistical estimator of $\phi$ has mean-square error which is bounded below by $1/\mathcal{I}(\phi)$.
We compute the Fisher information of our demonstration using two methods, both plotted in Fig.~\ref{FisherMonteML}. The first (solid black line) is directly computed using the experimentally extracted $p_i(\phi)$ in the relation $\mathcal{I}(\phi) = \sum_{i=1}^9 {p_i\left(\phi\right)}^{-1}\left({\textrm{d} p_i(\phi)}/{\textrm{d}\phi}\right)^2$, with error estimated using a Monte-Carlo simulation that assumes Poisson-distributed noise on the four-photon detection rates.
The second method is to obtain the variance $\Delta^2{\phi_j}$ of $\mathcal{M}$ maximum-likelihood estimates $\{\phi_j\}$, each using $\mathcal{N}$ photons, and evaluate the relation $\mathcal{I}_{\rm ML}=1/(\mathcal{N} \times \Delta^2{\phi_j})$. 
Note that maximum-likelihood estimation saturates the Cram\'{e}r-Rao bound and loses any bias as data is accumulated, and is 
practical 
for characterising an unknown phase when $p_i(\phi)$ are characterised.
We simulate $\mathcal{M}=10,000$ maximum-likelihood estimates for a discrete set of waveplate settings,
and for each estimate we sample $\mathcal{N}=1,000$ times from $p_i(\phi)$.
This number of samplings ensures unbiased and efficient estimation~\cite{br-jopa-25-3813}.
Computed values of $\mathcal{I}_{\rm ML}(\phi)$ are then plotted (circles) in Fig.~\ref{FisherMonteML}, showing close agreement with $\mathcal{I}(\phi)$.
}

\jmob{We also plot in Fig.~\ref{FisherMonteML} theory-predicted} Fisher information \jmob{computed from}
the POVM \jmob{description of our 
multiplexed detection system,}
taking into account the SPDC gain parameter $\tau$
and the total \jmob{circuit and detector} efficiency $\eta$ \jmob{of}
our setup. We find general agreement of the main features between theory and experiment \jmob{($\mathcal{I}$ and $\mathcal{I}_{\rm ML}$)}, while the discrepancy is attributed to imperfect waveplate rotations and imperfect temporal indistinguishability of multi-photon states. 

\pete{Fig.~\ref{FisherMonteML} also shows the shot-noise limit for two photons passing through the measured phase, computed on the basis of the average photon number in the sensing path. For our experiments $\tau<0.1$, which bounds the Fisher information for the target path and is computed
to lie in the range $2.01\pm0.01$.} The shaded region displays the quantum advantage over the shot-noise limit---the maximum advantage achieved in our experiment is 28.2$\pm$2.4\% at $\phi=3.91\pm0.06$ radians. The theoretical \pete{maximum} advantage that can be achieved by the scheme with our $\tau$ and $\eta$ parameters is 45$\%$.

An important feature of the theory and experiment curves in Fig.~\ref{FisherMonteML} is the  
troughs in $\mathcal{I}$
(similar features \jmob{were} presented elsewhere, e.g. the supplemental information for Ref.~\onlinecite{xi-natphot-5-43}), occurring about points where some or all of the fringes in Fig.~\ref{9probspatterns} have minima or maxima.  In contrast, when all decoherence processes and experimental imperfections are absent, $\mathcal{I}$
is predicted to be independent of phase rotation --- a common feature of metrology schemes using photon-number counting measurement \cite{ho-pra-79-033822}.  
The definition of 
$\mathcal{I}(\phi)$
reveals points of instability when the numerator 
$\textrm{d} p_i(\phi)/{\textrm{d}\phi}$
vanishes but $p_i\left(\phi\right)$ does not --- this will arise even with very-small experiment imperfections 
that lead to interference fringes with visibility $<1$.  
\jmob{A 
solution is to incorporate a reference phase} in conjunction with a feedback protocol to optimise estimation of an unknown phase~\cite{xi-natphot-5-43}.
The symmetry of the generalised singlet state at the heart of this scheme enables a control phase to be placed on the reference path as opposed to the sensing path in the traditional manner. 
We demonstrate the feasibility of the former by repeating our experiment
with a control phase rotation ($\theta$ in Fig.~\ref{setup1}) placed in the reference path $b$ that shifts the regions of maximal sensitivity with respect to the phase in the sensing path --- see Appendix for four-photon interference fringes and corresponding Fisher information. This may find practical application where the control phase has to be separated from the reference path. Furthermore, the reference path could be used for heralding to maximise the Fisher information per photon passing through the unknown sample using fast switching \cite{bo-prl-108-053601} of the sensing path conditioned on detection events at the reference path. Using heralding \jmob{and perfect photon-number resolving detection}, the entire downconversion state can achieve quantum advantage with the $\tau$ value from our experiment (see Appendix).

\jmob{We have demonstrated the key features of}
a promising technique for realising 
practical 
quantum-enhanced sensor\jmob{s}~\cite{ca-prl-105-013603}
\jmob{that are} robust to loss and \jmob{designed to use} 
a photon source 
\jm{based on}
current technology\jmob{, in contrast to other quantum technology schemes that rely on generating a fixed number of photons}. This now shifts the emphasis for practical quantum metrology onto \jmob{using}
low-loss circuitry and high-efficiency photon detection\jmob{;}
$93$\% efficient detectors operating in the infrared have recently been reported \cite{ma-nphoton-7-210}.
Natural extensions would be to implement the scheme in an integrated architecture with on-chip photon sources and detectors, thereby reducing optical loss and allowing for integration with micro-fluidic channels for bio-sensing \cite{cr-apl-100-233704}.
For a given efficiency $\eta$, the gain parameter $\tau$ in the down-conversion process also dictates the level of precision the scheme can achieve. As circuit loss is reduced, it would be beneficial to increase $\tau$ to the values ($\tau>$ 1) studied in Ref.~\onlinecite{ca-prl-105-013603}; enhancing SPDC with a cavity~\cite{kr-nphoton-4-170} may be a promising approach to achieve this.

\textbf{Acknowledgements.} The authors are grateful for financial support from, EPSRC, ERC, NSQI, NRF (SG), MOE (SG) and ARC CQC2T. 
JCFM is supported by a Leverhulme Trust Early Career Fellowship. 
GLP acknowledges support from the Benjamin Meaker Visiting Fellowship and from the ARC Future Fellowship.
JLOB acknowledges a Royal Society Wolfson Merit Award and a RAE Chair in Emerging Technologies.
%

\clearpage

\newpage

\section*{Appendix}

\subsection{Parametric Downconversion Setup.} 
Horizontal polarised 404nm pulsed light, generated by up-conversion of a Ti-Sapphire laser system (85fs pulse length, 80MHz repetition rate), is focused to a waist of 50$\mu$m within the crystal to ideally generate the state $\ket{PDC}$ at the intersection of the ordinary (o) and extra-ordinary (e) cones of photons \cite{kw-prl-75-4337} in paths $a'$ and $b'$ of Fig.~\ref{setup1}. Spatial and temporal walk-off between e and o light is compensated \cite{kw-prl-75-4337} with one half-waveplate (optic axis at $45^\circ$ to the vertical) and one 1mm thick BBO crystal in each of the two paths $a$ and $b$. The spectral width of ordinary and extraordinary light generated in type-II downconversion differs, leading to spectral correlation of the two polarisations. Setting one waveplate to $90^\circ$ and aligning the two paths onto a PBS separates the e and o light, sending all e light onto output $a$ and all o light onto output $b$\cite{ki-pra-67-010301}. This removes spectral-path correlation in the PDC state, leaving only polarisation entanglement across paths $a$ and $b$, and thus erasing polarisation dependent loss in the sensing path and the reference path of the setup. 

\subsection{Number resolving photon detection.}
We approximate number resolving detection using a multiplexed method \cite{xi-natphot-5-43}. Photons in each of the four modes $a_h$, $a_v$, $b_h$ and $b_v$ are symmetrically distributed across $d=4$ detector modes using one-to-four \jmob{optical} fibre splitters, and photons are detected at each of the outputs using a total of 16 silicon Avalanche Photodiode ``bucket'' Detectors (APDs). Each detector has two possible outcomes: no detection event (``0'') for a vacuum projection and a detection event (``1'') for detection of one-or-more photons, with nominal $\sim60\%$ efficiency.

We constructed a coincidence counting system based around a commercial 
time-correlated single photon counting (TCSPC) system. This system time-tags incoming photons across sixteen channels with $\sim$80ps timing resolution, and logs the timetags on a PC. We developed fast routines, running on a CPU, which efficiently count, store and display instances of every possible $N$-photon coincidence pattern (up to $N$=16---65,536 possible patterns) using these timetags, in real-time. We then compute photon number statistics from these coincidence count-rates. 

\subsection{Derivation of the POVM operators for approximate photon counting using multiplexed arrays of single-photon detectors.}

To implement approximate photon counting, we use 
a series of fibre splitters
to distribute incoming photons 
across $d$ single-photon detectors  (here $d=4$).  The ratios of the fibre splitters 
are chosen to implement a unitary transformation, 
denoted by $U_{\rm mp}$, which implements the following transformation according to the
relation on the mode annihilation operators $m_j$:
\begin{eqnarray}
m_{j}^{\rm out}&=&U_{\rm mp}^{\dag}m_{j}^{\rm in}U_{\rm mp}\nonumber\\
&=&(1/\sqrt{d})m_1^{\rm in}+ \sum_{j=2}^d s_j m_j^{\rm in}
\label{eq:fanouttransformation}
\end{eqnarray}
for arbitrary scalars $s_j$.

A single-photon detector is described by a POVM with elements $\vert0\rangle\!\langle0\vert$ and $\sum_{c=1}^{\infty}\vert c\rangle\!\langle c\vert$ in the Fock basis, corresponding to 0 or 1 detection event(s) respectively.  Mode $m_1$ labels one of the principle modes from the experiment, from the set $\{a_h,a_v,b_h,b_v\}$, and $m_2,\cdots,m_{d}$ are ancillary modes, initially in the vacuum state.  Two standard methods for implementing the required transformation $U_{\rm mp}$ are: (i) A sequence of splitters first on pair $\{m_1,m_2\}$ with transmissivity $1/d$, then on pair $\{m_2,m_3\}$ with transmissivity $1/(d-1)$, and so on, finishing with a $50:50$ splitting of pair $\{m_{d-1},m_d\}$. (ii) A tree of $50:50$ spitters.  Method (i) works for arbitrary numbers of detectors, whereas (ii) is suitable only when the number of detectors is a power of two.  The current experiment implements (ii) for multiplexing four detectors.  Eq.~(\ref{eq:fanouttransformation}) is easily verified for both (i) and (ii).

\begin{figure}
\includegraphics[width=
\columnwidth]{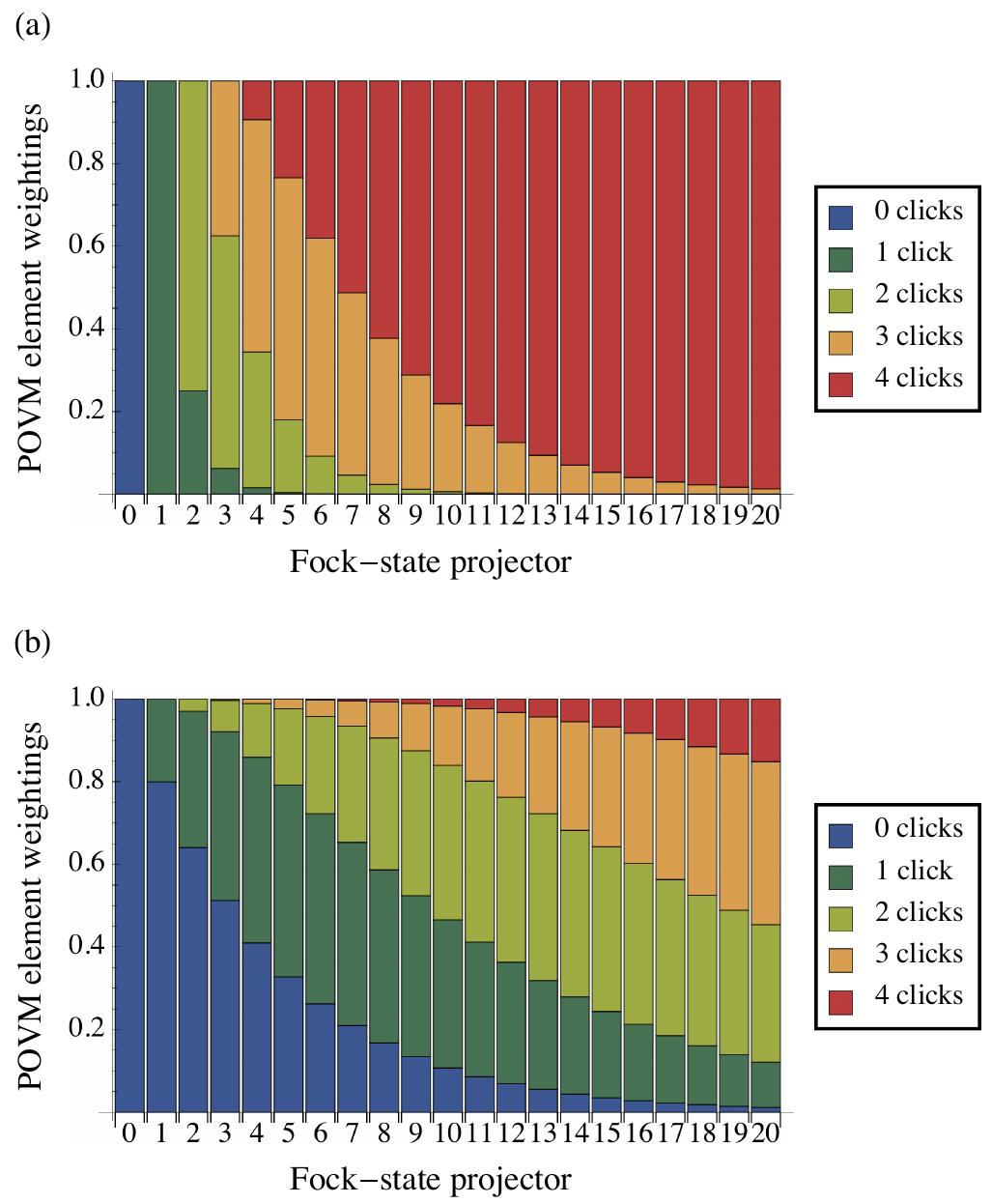}
\vspace{-0.5cm}
\caption{
\textbf{POVM elements for photon counting $n$ using four multiplexed APDs 
}
(a) $\eta=1$;
(b) $\eta=0.2$
}
\label{weightingsFig}
\end{figure}

Suppose now 
$r$ detection events are registered at the first $r$ detectors, and the remaining $d-r$ detectors do not detection event.  The probability of this event for an arbitrary input state $\rho_{\rm in}$ incident in mode $m_1$ is: 
\begin{eqnarray*}
P&=&{\rm tr}\Bigg[ \bigotimes_{j=1}^r  \left( \sum_{c_j=1}^\infty \vert c_{j}\rangle \! \langle c_j\vert \right)\otimes
 \vert {\bf 0}\rangle \! \langle {\bf 0} \vert _{r+1\cdots d}.\\
&&  \qquad \qquad .\left(
 U_{\rm mp} \rho _{\rm in}\otimes
 \vert {\bf 0}\rangle \! \langle {\bf 0} \vert_{2 \cdots d} U_{\rm mp}^{\dag }
 \right)
 \Bigg] \nonumber \\
&=&\sum_{c_1\cdots c_r=1}^\infty\langle c_1\cdots c_r {\bf 0}\vert
\big(
U_{\rm mp}\rho _{\rm in}\\
&& \qquad \qquad \otimes \vert {\bf 0}\rangle \! \langle {\bf 0}\vert _{2\cdots d}U_{\rm mp}^{\dag }
\Big)
\left\vert c_1\cdots c_j {\bf 0}\right\rangle _{1\cdots d} \nonumber \\
&=&\sum_{c_1\cdots c_r=1}^\infty\frac{1}{c_1\cdots c_r}\frac{1}{d^c}\langle 0\vert m_1^c\rho _{\rm in} m_1^{\dag c}\left\vert 0\right\rangle_1,
\end{eqnarray*}
where
$c=\sum_{j=1}^r c_j$.  Taking into account that the $r$ detection events can occur in $d \choose r$ equivalent configurations, the complete POVM element corresponding to $r$ detection events across the multiplexed detector is given by:
\begin{equation}
E_r={d \choose r}\sum_{c=r}^\infty\frac{1}{d^{c}}\left( \sum_{\{c_j|c_j\mbox{ sum to }c\}}\frac{c!}{c_1!\cdots c_r!}\right)
\vert c\rangle \!\langle c\vert. \nonumber
\end{equation}
The combinatorial quantity $\frac{1}{r!}\sum_{\{c_{j}|c_{j}\mbox{ sum to }c\}}\frac{c!}{c_1\cdots c_r}$ is the same as the Stirling number of the second kind, denoted $S(c,r)$, which counts the number of ways of partitioning $c$ objects into $r$ {\it non-empty} subsets.  Finally then,
\begin{equation*}
E_r=\sum_{c=r}^{\infty }w_r\left( c\right) \vert c \rangle \langle c \vert
\end{equation*}
where,
\begin{equation}
\label{eq:fanoutweights}
w_r\left(c\right)=\frac{d! S(c,r)}{(d-r)! d^c}
\end{equation}
as derived in\cite{sp-pra-85-023820} using a different method. These weights are illustrated in Fig.~\ref{weightingsFig} (a) for the case $d=4$.  This result can also be verified inductively 
using standard relations for the Stirling numbers.   
The 
bound $w_r(c)\leq S(c,r)/d^{c-r}$ implies that, for $c>r$, $w_{r}(c)\longrightarrow 0$ as $d\longrightarrow\infty$, and the completeness property of  POVM 
implies that $w_r(r)\longrightarrow1$ as $d\longrightarrow\infty$.  In other words, the POVM element corresponding to $r$ detection events converges to the perfect projector $\vert r \rangle\langle r \vert$ in the large $d$ limit, as expected.

\subsection{Photon losses.} To incorporate photon losses in our analysis, we use
a standard loss model for which the mode in question is coupled via a beamsplitter to an ancillary mode, initially in the vacuum state, which is traced out at the end.  
We assume that losses are polarization independent and all
single-photon detectors in a multiplexed array are modelled with the same efficiencies $\eta_d$,
and hence detector loss can be incorporated 
as a loss channel with efficiency $\eta_d$ to the combined POVM Eq.~(\ref{eq:fanoutweights});
this loss commutes with fiber splitters and can be considered as part of the combined system efficiency.
The effect of system efficiency $\eta$ can be incorporated into the multiplexed POVM by the linear transformation:
\begin{equation}
\left\vert c\right\rangle \left\langle c\right\vert \mapsto \sum_{c'=c}^{\infty }{c' \choose c}\eta^c(1-\eta )^{c'-c}\vert c' \rangle\langle c' \vert.
\end{equation}

It is important to remove polarisation-dependent loss in order to preserve symmetry
properties of the downconversion state, Eq.~(\ref{SPDCsinglet}), which follow from it being
a superposition of singlet states, namely: $U \otimes U \ket{PDC}=\ket{PDC}$,
where $U$ is an arbitrpary unitary rotation of two polarization modes.  This symmetry implies
a simple structure for the mixed downconversion state which arises after the effects of photon losses;
$\rho_{PDC}=\Sigma_{n_a,n_b} P(n_a,n_b) \rho_{n_a,n_b} $, where $n_{a(b)}$ denotes the total photons
across the $a(b)$ modes, and the transformation implemented by a unitary polarization rotation acts
independently on the ($n_a,n_b$) subspaces.  For the diagonal subspaces ($n_a=n_b$),
$\rho_{n_a,n_a}$ is a mixture of a singlet state with $2n_a$ photons, together with decoherence terms. The weights $w_r(c)$ are altered correspondingly as illustrated in Fig.~\ref{weightingsFig} (b).

\subsection{Characterising gain $\tau$ and efficiency $\eta_b$, $\eta_a$} 

The probability of generating one pair of photons in SPDC is computed via Eq.~(\ref{SPDCsinglet}) is given by
\begin{eqnarray}
p= \frac{2 \tanh^2(\tau)}{\cosh^4(\tau)}
\end{eqnarray}
All the following are constant with respect to the phase rotation $\phi$ and can be taken as the $\phi$-average over experimental data. Summing pairs of proper singles (single photons that are detected and not part of a coincidence event with $\geq1$ other photon event/ with no events elsewhere in the detection scheme):
\begin{eqnarray}
P(1,0,0,0)+P(0,1,0,0)=p\eta_b(1-\eta_a)\\
P(0,0,1,0)+P(0,0,0,1)=p(1-\eta_b)\eta_a
\end{eqnarray}
Summing all four two-fold coincidences yields
\begin{eqnarray}
&&P(1,0,0,1)+P(0,1,1,0)+P(1,0,1,0)+P(0,1,0,1)\nonumber\\
&&=p\eta_b\eta_a
\end{eqnarray}
This is then be solved for $\eta_b$ and $\eta_a$, then for $p$, and hence $\tau$ via a cubic equation.
\vspace{1cm}

\subsection{Shifting the output interference fringes with a control phase}
\begin{figure}[b!]
\includegraphics[width=7.5cm]{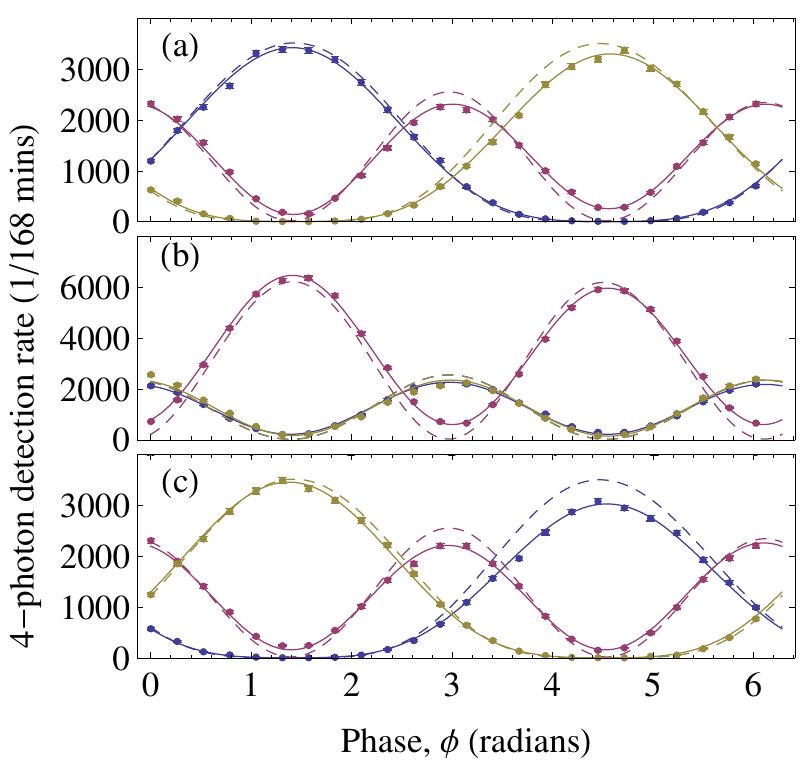}
\vspace{-0.cm}
\caption{
\textbf{Shifted four-photon detection patterns} (circular points) plotted as a function of phase $\phi$ (equivalently, waveplate angle $\phi/4$) for $\tau=0.055$, $\eta_{as}=0.24$ $\eta_{b}=0.13$.
The theory and measured detection patterns and Poisson error is plotted in the same manner as Fig.~\ref{9probspatterns} of the main text.
}
\label{9probspatterns23rd}
\end{figure}
\begin{figure}[b!]
\includegraphics[width=\columnwidth]{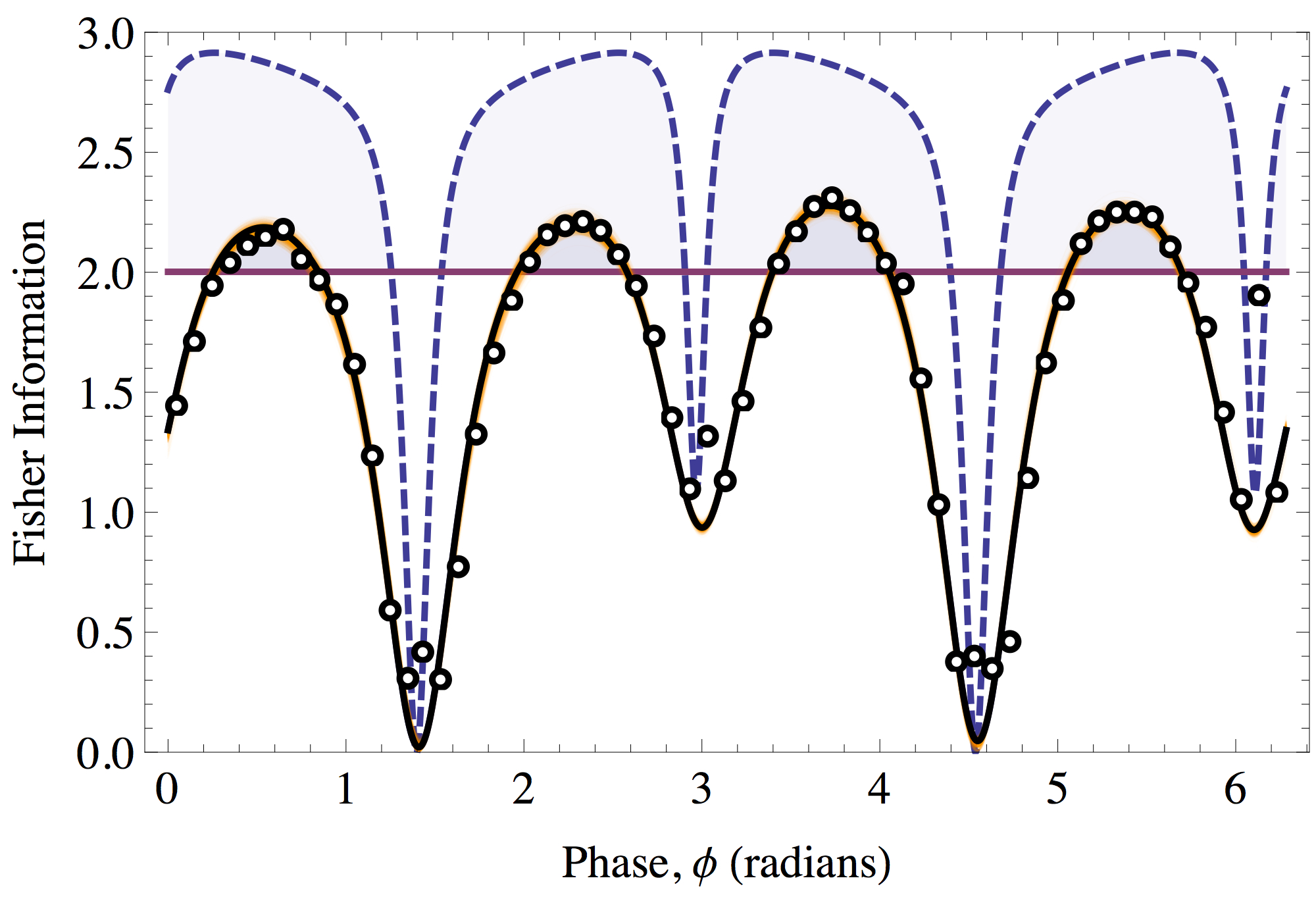}
\vspace{-0.cm}
\caption{
\textbf{Shifted Fisher Information extracted from Fig.~\ref{9probspatterns23rd}.}
Curves and data points are plotted in the same manner as Fig.~\ref{FisherMonteML} of the main text, with theoretical Fisher information computed with the parameters
$\tau=0.055$, $\eta_{a}=0.24$ $\eta_{b}=0.13$.
The maximum experimentally derived Fisher information $\mathcal{I}$ surpasses the shot-noise limit by $13.88\pm0.95\%$ at phase $\phi=3.71\pm0.01$ radians.
}
\label{FisherMonteML23rd}
\end{figure}

Due to the phase dependence of precision (Fisher Information) in many metrology schemes, it is desirable to maximise the value of precision for a given scheme using a control phase. Typically, this is performed using a control phase inside an interferometer or a sequential interferometer in the same beam-path as the path used for direct sensing. The symmetry of the singlet state used in the scheme demonstrated here enables the control phase to be moved onto the reference path. We demonstrate this by using the control waveplate $\theta$ in Fig.~\ref{setup1} of the main text to shift the interference patterns (Fig.~\ref{9probspatterns23rd}), and therefore the Fisher information (Fig.~\ref{FisherMonteML23rd}), by 20 degrees. Note that the Fisher Information plots retain the same periodic structure of Fig.~\ref{FisherMonteML} of the main text, as expected.

\subsection{Fisher Information attainable from heralding}\label{HeradingAppendixSection}
Using Type-II SPDC quantum metrology~\cite{ca-prl-105-013603}  has the benefit of correlations across the sensing path $a$ with the interferometer and a reference path $b$, which could be used with heralding and subsequent gating on the sensing path to optimise further precision of the scheme. Tables~I~and~II show the computed Fisher information obtainable in principle in the scheme demonstrated with $\tau$ similar to what we have in our experiment and with the inclusion of heralding and fast switching to act as a gate to optimise photon flux through an unknown phase. We have assumed $d\rightarrow \infty$ for the multiplexed photon detection setup and for simplicity the total circuit and detector efficiency $\eta$ is the same across all four modes $a_h, a_v, b_h, b_v$. The Fisher information is computed for the detection outcomes at the output of the sensing path $a_h,a_v$, conditional on detecting $K$ photons in any pattern across the output of the reference path $b_h, b_v$.

\begin{table}[h!]
\caption{Fisher information per photon for $\tau = 0.05$, and the $\eta$ in the table, conditional on heralding $K$ photons in the reference path.
}
\centering
\begin{tabular}{c c c c c c}
\hline\hline
\begin{tabular}{c}
Herald\\ photons, $K$
\end{tabular} & $\eta=$0.7 & $\eta=$0.8 & $\eta=$0.9 & $\eta=$0.95 & $\eta=$1 \\ [0.5ex] 
\hline
0 &	0.48994 &	0.64043 &	0.81125 &	0.90431 &	1.0025 \\
1 &	0.69835 &	0.79934 &	0.90071 &	0.95155 &	1.0025 \\
2 &	0.79119 &	0.95866 &	1.13993 &	1.23574 &	1.335 \\
3 &	0.85610 &	1.08952 &	1.35927 &	1.50862 &	1.66806 \\ [1ex]
\hline
\end{tabular}
\label{Tau_0_05}
\end{table}
\begin{table}[h!]
\caption{Fisher information per photon for $\tau = 0.1$, and the $\eta$ in the table, conditional on heralding $K$ photons in the reference path.}
\centering
\begin{tabular}{c c c c c c}
\hline\hline
\begin{tabular}{c}
Herald\\ photons, $K$
\end{tabular}
 & $\eta=$0.7 & $\eta=$0.8 & $\eta=$0.9 & $\eta=$0.95 & $\eta=$1 \\ [0.5ex] 
\hline
0 &	0.48964 &	0.64159 &	0.81493 &	0.90974 &	1.01003 \\
1 &	0.69331 &	0.79726 &	0.90281 &	0.95621 &	1.01003 \\
2 &	0.78482 &	0.95468 &	1.13974 &	1.238 &	1.34004 \\
3 &	0.84795 &	1.08358 &	1.35745 &	1.50959 &	1.67226 \\ [1ex]
\hline
\end{tabular}
\label{Tau_0_1}
\end{table}

\end{document}